\documentclass[conference]{IEEEtran}
\IEEEoverridecommandlockouts
\usepackage{cite}
\usepackage{amsmath,amssymb,amsfonts}
\usepackage{algorithmicx}
\usepackage{textcomp}
\usepackage{xcolor}
\usepackage{algorithm}
\usepackage{algpseudocode}
\usepackage{graphicx,subfigure}
\usepackage{fancyhdr}  
\usepackage{cases}
\usepackage{extarrows}
\usepackage{multirow,tabularx}
\usepackage{mathtools}
\usepackage[english]{babel}
\usepackage{float}
\usepackage{subfig}
\def\BibTeX{{\rm B\kern-.05em{\sc i\kern-.025em b}\kern-.08em
		T\kern-.1667em\lower.7ex\hbox{E}\kern-.125emX}}

\begin{document}
\title{Joint Beamforming and Phase Optimization in a Multi-user Communication System Composed of Dual Reconfigurable Intelligent Surfaces}

\author{
	\IEEEauthorblockN{
		RuiTianYi~Lu\IEEEauthorrefmark{1}
		}
	\IEEEauthorblockA{
		\IEEEauthorrefmark{1}School of Information Engineering, Wuhan University of Technology, Wuhan 430070, China \\
		\texttt{lrty@whut.edu.cn}}
}
\maketitle

\begin{abstract}
Reconfigurable Intelligent Surface (RIS) has becoming a useful tool in future wireless communication systems for close-distance communication network. This paper we use Reconfigurable Intelligent Surface (RIS) for downlink multi-user communication designed to improve energy collection performance while satisfying wireless information and Power Transfer (WIPT). The designed system consists of an IRS-assisted system consists of a multi-antenna assisted base station (BS) and two opposite multi-antenna assisted information receiver cooperated (RIS) as energy receiver (ERs) that meets energy collection requirements. Based on the electromagnetic property of Reconfigurable Intelligent Surface (RIS), like two mirrors that are opposite each other, setting two Reconfigurable Intelligent Surface (RIS) attached to the city buildings to reflect the sending signals. The transmitting precoding of the Multi-antenna Auxiliary Base Station (BS) and the angular phase transfer matrix of the multi-antenna Auxiliary Information Receiver (IRs) need to be optimized together to maximize the energy harvesting of IoT devices for energy efficiency (EE) of the IRs system and to provide users with the efficiency of the received signal. In order to solve the joint optimization problem effectively, we turn the non-convex maximize problem into the equivalent formal error method based on the mean square, and finally use the iterative algorithm for optimization. As for algorithm, we respectively use MSE method, semidefinite relaxation techniques to simplify transmitting beamforming matrix and the matrix phase shift. Through the observation of simulation data, it can be concluded that the performance optimization method of SDR based on RIS is effective.
\end{abstract}

\begin{IEEEkeywords}
Reconfigurable intelligent surfaces,MISO, energy efficiency, phase shift, non-convex optimization, alternating maximization, gradient descent, sequential fractional programming.
\end{IEEEkeywords}

\section{Introduction}\label{sec:intro}
It is the fifth-generation (5G) mobile communications that appeals to the world's attention in 2020, including massive multiple-input multiple-output(MIMO), millimeter wave (mmWave) communications, and ultra-dense networks \cite{Larsson2014Massive,Yong2010MIMO,Palomar2003Joint,Zhang2013MIMO,Lu2014An}. The problem of spectrum shortage is dealt with by millimeter wave (mmWave). Admittedly, 5G has brought a lot of changes to our life, such as virtual reality, autonomous driving, uav, etc. 6G can be roughly considered as an extension of 5G . In the upcoming 6G communication\cite{Chowdhury20206G}, we are required to solve the massive energy loss caused by numerous cellular network base stations (BSs) \cite{Zhang2019Energy} and the hardware cost caused by the above technologies of substantial power-hungry radio-frequency (RF) chains regained in MIMO/mmWave systems \cite{Mehta2005RF}. To solve the above problems, reflecting intelligent surface (RIS) surface on the physical layer is adopted to improve the energy efficiency of wireless communication \cite{Ye2020Joint,Yan2020Passive}. The behavior entity between transmitter and receiver will reduce the quality of the received signal. This is due to the uncontrollable interaction of the radio waves emitted with objects around them. However, the emergence of reconfigurable intelligence surfaces in wireless communications makes it possible for operators to overcome the negative effects of natural wireless propagation by controlling the scattering, reflection, and refraction characteristics of radio waves. Recent research results show that it is a reconfigurable intelligent surface that can effectively control wavefront, such as phase, amplitude, frequency, etc., without complex decoding, encoding, and polarization processing operations of RF impact signals.

RIS is a flexible electromagnetic transmission concept, consisting of a large number of square metal blocks programmable flat cell matrix, whose each unit can be passively independent digital controlled by external signals real-time to manipulate the reflected wave electromagnetic response, such as phase and amplitude\cite{Basar2020Indoor,Mursia2020RISMA}. Functions at the physical level are software-defined, and given the software-defined interaction between intelligent surfaces and incident waves, the term software-defined surface (SDS) can also be used for these intelligent surfaces\cite{Sheng2020P}. In other words, since the reflective properties of these smart surfaces/walls/arrays in the physical layer can be controlled by software, they can be referred to as SDS. The appearance of intelligent reflective surface satisfies the characteristic of 6G short wave length and has the practicability of being used in high-rise buildings. Most importantly, the ability to encode wireless communication EM carrier signals allows a wireless communication system based on reconfigurable intelligent surface to directly modulate the electromagnetic carrier signals, instead of the traditional RF chains \cite{Huang2018Reconfigurable}. Therefore, it has great potential for realizing UM-MIMO and holographic MIMO technologies \cite{Palomar2003Joint,Zhang2013MIMO}. In order to study the specific application of RIS, we assume that the channel state information CSI is known. Existing research work includes channel estimation, combined passive beamforming, IRS phase shift matrix optimization, and BS transmission beamforming optimization. Different from traditional precoding, which only precodes at the emitter, we consider the combined optimization of emitter beamforming and IRS phase shift matrix to optimize the transmission characteristics of RIS system.

As an emerging technology, RIS is faced with much chanllenges, which brings out our model design. In terms of the transmission pattern firstly, a certain amount of energy is required to excite the RIS for transmission. However, due to the limited computing and bandwidth resources of the base station, the energy emitted by the base station is limited by the maximum power. Secondly, how to adjust the RIS reflection phase is the main research target for the beam energy transfer matrix emitted by the base station. Because RIS does not change amplitude, only phase. 

This paper is aimed to explore the feasibility of high order modulation for MIMO wireless transmission by RIS \cite{Tang2020Design} and to improve the transmission efficiency of the wireless communication system by analyzing and modeling the experimental results of RIS-aid wireless system. Specifically, we consider a user communication system consisting of a multi-antenna base station and two cooperated reconfigurable intelligent surfaces. Because there is a significant propagation loss from the base station directly to cell edge users, we deploy RIS to help base station serve multiple cell edge users. It is practical that RIS can be attached to buildings to provide two  different high BS-IRS links and two different IRS-USER links to establish the probability of line-of-sight (LoS) propagation. The two RIS will have different energy transfer matrices from the base station and transfer matrix to users. In order to optimize the system, they have different phase modulation matrices thus leading to different transmit beamforming matrices from base station which are completely independent of each other. Multiple antennas in the base station transmit different frequency waves, which are reflected by the RIS to the users receiving the specific frequency respectively. The information received by each cell-Edge user is observed in Figure
1, assuming no direct base station propagation and all derived from RIS reflection. By carefully accommodating RIS phase shift, multi-user interference in the system can be further suppressed, and the energy loss from direct base station propagation is reduced. In this paper, we focus on optimizing the total signal noise ratio of multiple users. Based on convex optimization problem low complexity sub-optimal resource allocation algorithms are proposed. The simulation results demonstrate the advantages of our proposed RIS-Aided system, which can help us optimize the delivery efficiency. The main contributions of this paper are summarized as follows:

We investigated a multi-user, multi-unit network consisting of a multi-antenna base station BSs and an RIS, which is used as a mediator to enhance data transmission to users. In order to take quantitative assessment, the goal of this paper is to maximize the EE maximization problem by jointly optimizing the transmit beamforming matrix at the BSs and the phase shift matrix at the RIS.

The nonconvex optimization problem of objective function design in this paper is difficult to solve optimally. Therefore, we first use the equivalent accessibility problem method of mean square error (MSE) \cite{Murphy1996General,Jinyong2010Estimation} to transform the objective function problem. Having stable base station transmits energy power, what makes this research unique is how to allocate two transmitting power paths under two different RIS circumstances. As for multi-user system, in practice, we will transfer for fixed IRS phase shift matrix beamforming into a second-order cone programming, for the launch of a fixed beamforming matrix, the IRS phase offset matrix is optimized based on the semidefinite relaxation (SDR) technique\cite{Luo2010Semidefinite}. Finally, iterative alternate optimization algorithm is also used to obtain the transmitted beamforming matrix and RIS phase shift matrix.

The results of a large number of simulation data show that significant throughput gain can be achieved with the help of RIS, and the transmission efficiency of wireless communication system can be significantly improved by using the algorithm designed in this paper. The remainder of this paper is organized as follows: 

Related works are introduced in Section II. Section III illustrates the system model. Optimization formulation and analysis is simplified in Section IV. The simulation data results are shown in Section V. Section VI concludes the paper and talk about future work.

\textit{Notation}: $a$ is a scalar, $\mathbf{a}$ is a vector, and $\mathbf{A}$  is a matrix. $\mathbf{A}^T$, $\mathbf{A}^H$, $\mathbf{A}^{-1}$, $\mathbf{A^+}$, and $\|\mathbf{A}\|_F$ denote transpose, Hermitian (conjugate transpose), inverse, pseudo-inverse, and Frobenius norm of $ \mathbf{A} $, respectively. $\mathrm{Re}(\cdot)$, $\mathrm{Im}(\cdot)$, $|\cdot|$, $(\cdot)^*$ and $\mathrm{arg}(\cdot)$ denote the real part, imaginary part, modulus, conjugate and the angle of a complex number, respectively. $\text{tr}(\cdot)$ denotes the trace of a matrix and $\mathbf{I}_n$ (with $n\geq2$) is the $n\times n$ identity matrix. $ \mathbf{A} \circ \mathbf{B} $ and $ \mathbf{A} \otimes \mathbf{B} $ denote the Hadamard and Kronecker products of $\mathbf{A}$ and $\mathbf{B}$, respectively, while $\text{vec}(\mathbf{A})$ is a vector stacking all the columns of $\mathbf{A}$. $ \mathrm{diag}(\mathbf{a})$ is a diagonal matrix with the entries of $\mathbf{a}$ on its main diagonal. $\mathbf{A}\succeq\mathbf{B} $ means that $\mathbf{A}-\mathbf{B}$ is positive semidefinite. Notation $x\sim\mathcal{CN}(0,\sigma^2)$ means that random variable $x$ is complex circularly symmetric Gaussian with zero mean and variance $\sigma^2$, whereas $E[x]$ denotes $x$'s expected value. $\mathbb{R}$ and $\mathbb{C}$ denote the complex and real number sets, respectively, and $j\triangleq\sqrt{-1}$ is the imaginary unit.

\section{Related Works}\label{sec:format}

Based on Reinforcement Learning we also can conduct Intelligent Reflecting Surface Assisted Anti-Jamming Communications \cite{Yang2020Intelligent}. It has become one of the important challenges for wireless communication security that intelligent jammer can launch malicious jamming to attack legitimate transmission. Therefore, this paper uses the intelligent reflector surface (IRS) to improve the anti-jamming communication performance and reduce the jamming interference by adjusting the IRS surface reflection element. In order to improve the communication performance of intelligent jammer, a joint optimization problem of base station power allocation and IRS reflected beamforming is proposed. As the jamming model and jamming behavior are dynamic and unknown, a fast policy hill-climbing (Wolf-PHC) \cite{Xi2015A} learning approach is proposed to jointly optimize the anti-jamming power allocation and reflecting beamforming strategy without the knowledge of the jamming model.

NOMA network can be aided to downlink analysis for Reconfigurable Intelligent Surfaces \cite{Yiqing2020Joint}. By activating blocked users and altering successive interference cancellation (SIC) sequences, reconfigurable intelligent surfaces (RISs) become promising for enhancing nonorthogonal multiple access (NOMA) systems \cite{Benjebbour2013Concept}. This work investigates the downlink performance of RIS-aided NOMA networks via stochastic geometry. They first introduce the unique path loss model for RIS reflecting channels. Then, they evaluate the angle distributions based on a Poisson cluster process (PCP) framework, which theoretically demonstrates that the angles of incidence and reflection are uniformly distributed. Lastly, they derive closed-form expressions for coverage probabilities of the paired NOMA users. It can be proved that 1) RIS-aided NOMA networks perform better than the traditional NOMA networks; and 2) the SIC order in NOMA systems can be altered since RISs are able to change the channel gains of NOMA users\cite{Liu2018Performance}.
 
Reconfigurable Intelligent Surfaces not also can carry the signal from base station to users, there are plenty of papers discussing multiuser full-duplex two-way communications via Intelligent Reflecting Surface \cite{Zhang2020Sum,Sharma2020Intelligent}. Low-cost Passive Intelligent Reflector Surface (IRS) has recently been heralded as a revolutionary technology that can reconfigure the wireless propagation environment by carefully adjusting the reflector elements. An IRS scheme covering cellular multi-user full-duplex (FD) \cite{Akcapinar2015Full} bidirectional communication link dead zones while suppressing user-side self-interference and co-channel interference (CI) is proposed. This method allows the base station (BS) and all users to exchange information simultaneously, potentially doubling the spectral efficiency. To ensure network fairness, they jointly optimized the precoding matrix of BS and the reflection coefficient of IRS to maximize the weighted minimum rate (WMR) for all users under the constraints of maximum transmitted power and unit mode. They reexpress the non-convex problem and decouple it into two subproblems. Then, block coordinate descent (BCD) algorithm \cite{Qin2013Efficient} is used to conduct alternate optimization of the optimization variables in the equivalent problem. To further reduce computational complexity, they proposed a minimum-maximization (MM) algorithm to optimize the precoding matrix and reflection coefficient vector by defining the minimization function in the agent problem. Finally, simulation results confirm the convergence and efficiency of their proposed algorithm, and validate the advantages of introducing IRS to improve coverage in blind areas.

Besides deploying on wireless communication, Reconfigurable Intelligent Surfaces can play an important part in unmanned aerial vehicle (UAV) communication \cite{2020Reconfigurable}. This article \cite{2016Intelligent} quotes the investigation of the problem of maximizing the security and energy efficiency of a reconfigurable intelligent surface (RIS) assisted upline wireless communication system, in which the UAV is equipped with an RIS work between a mobile relay base station (BS) and a group of users. Their focus is on maximizing the system's safe energy efficiency by jointly optimizing the drone's trajectory, the RIS's phase shift, user association and transmit power. To tackle this problem, they divide the original problem into three sub-problems, and propose an efficient iterative algorithm. In particular, the successive convex approximation (SCA) method \cite{Tao2012Successive} is applied to solve the nonconvex UAV trajectory \cite{Wu2018Joint}, the RIS's phase shift, and transmit power optimization sub-problems. They further provide two schemes to simplify the solution of phase and trajectory sub-problem. Simulation results demonstrate that the proposed algorithm converges fast, and the proposed design can enhance the secure energy efficiency by up to 38$\%$ gains, as compared to the traditional schemes without any RIS.

Above this we browsed many aspects of RIS applications, I also made notes on RIS' own physical electromagnetic properties. This paper writes about analytical modeling of the path-loss for Reconfigurable Intelligent Surfaces-anomalous mirror or scatterer \cite{unknown}. A fundamental component for analyzing and optimizing RIS-empowered wireless networks is the development of simple but sufficiently accurate models for the power scattered by an RIS \cite{2011Light}. By leveraging the general scalar theory of diffraction and the Huygens-Fresnel principle \cite{2016Perfect}, they introduce simple formulas for the electric field scattered by an RIS that is modeled as a sheet of electromagnetic material of negligible thickness. The proposed approach allows them to identify the conditions under which an RIS of finite size can or cannot be approximated as an anomalous mirror.

\section{System Model}\label{sec:format}
In this section, we describe the system model consisting of two RIS and single base station for multiple users. The base station supports constant power value for two independent transmission channels, via RIS phase modulation respectively, reaching users. By solving the joint design of coupling transmitted powers and RIS phase shifts, we can maximize total energy efficiency of users.
\subsection{System Model}
\begin{figure} \vspace{-1mm}
  \begin{center}
  \includegraphics[width=90mm]{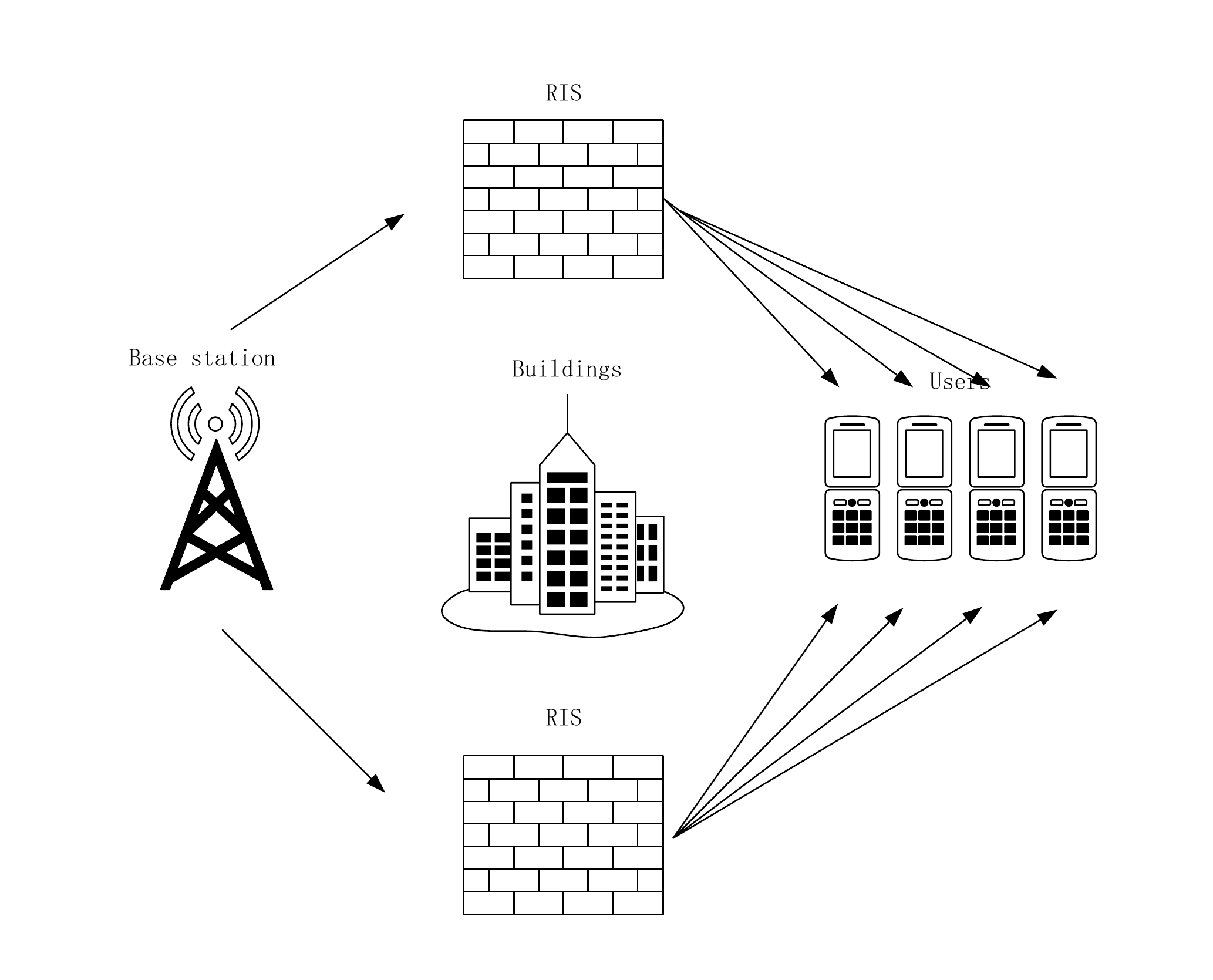}  \vspace{-2mm}
  \caption{The considered RIS-based multi-user MIMO system comprising of a $M$-antenna base station simultaneously serving in the downlink $K$ single-antenna users. RIS is assumed to be attached to a surrounding building's facade, and the transmit signal propagates to the users via the assistance of RIS that is capable of reconfigurable behavior.}
  \label{fig:Estimation_Scheme} \vspace{-6mm}
  \end{center}
\end{figure}

Consider a downlink communication system consisting of one BS equipped with M antenna element and K single-antenna mobile users. Let ${v} \in\mathbb{C}^{M \times K}$ and $s$ represent the transmit beamforming vector from BS to user and the data symbol for users with unit energy, respectively. We assume that this communication takes place via two RIS with N reflecting elements deployed on the facade of a building existing in the vicinity of both communication ends, as illustrated in Fig. 1. The direct signal path between the BS and the mobile users is intercepted due to the buildings between them.

The received signal at the $k$-th user is given by
\begin{equation}
\begin{aligned}
y_k = g_{1k} \mathbf{\Phi_1}\mathbf{H_1} v_{1k} s_k
        + g_{2k} \mathbf{\Phi_2}\mathbf{H_2} v_{2k} s_k + \\\sum^K_{\bar{k} = 1,\neq k}  ( g_{1k} \mathbf{\Phi_1}\mathbf{H_1} v_{\bar{1k}} + g_{2k} \mathbf{\Phi_2}\mathbf{H_2} v_{\bar{2k}}) + \sigma^2
\end{aligned}
\end{equation}
where $g_{k} \in  \mathbb{C}^{K \times N}$ suggests the receiving matrix between RIS and $k$-th user, $\mathbf{\Phi}\in \mathbb{C}^{N \times N} $ represents the phase shift matrix of RIS, $\mathbf{H} \in \mathbb{C}^{N \times M} $ represents the transmission characteristics from the base station to the RIS.

Spectrum efficiency
\begin{equation}
\begin{aligned}
&C_k = \log_2\\&\left(1+
\frac{| g_{1k} \mathbf{\Phi_1}\mathbf{H_1} v_{1k}|^2 + |g_{2k} \mathbf{\Phi_2}\mathbf{H_2} v_{2k}|^2 }
{\sum^K_{\bar{k}=1,\neq k}  (| g_{1k} \mathbf{\Phi_1}\mathbf{H_1} v_{\bar{1k}}|^2 + |g_{2k} \mathbf{\Phi_2}\mathbf{H_2} v_{\bar{2k}}|^2) + \sigma^2} \right)
\end{aligned}
\end{equation}

For convenience, let us define the equivalent channel spanning from the BS to the $k$-th user $\overline{h}_{1k} = g_{1k} \mathbf{\Phi_1}\mathbf{H_1}$ and $\overline{h}_{2k} = g_{2k} \mathbf{\Phi_2}\mathbf{H_2}$.

\subsection{Total Power Consumption Model}\label{sec:problem}
Composed of the BS transmit power, the hardware static power consumed in the BS, mobile user terminals and RIS, $k$-th user will consume the power $\mathcal{P}_{k}$ to connect with BS, which can be expressed as
\begin{align}\label{power_model1}
\mathcal{P}_k\triangleq \beta p_k + P_{{\rm U},k}+ P_{\rm B} + P_{\rm R},
\end{align}
where $\beta \triangleq \nu^{-1}$, $\nu$ is the efficiency of the transmit power amplifier, $p_k$ represents the transmit power,  $P_{{\rm U},k}$ denotes the hardware static power of $k$-th UE, $P_{\rm B}$ and $P_{\rm R}$ denote the total hardware static power consumption by BS and RIS, respectively.
The power consumption of RIS with $N$  reflecting elements which have the same energy dissipation can be written as $ P_R = NP_n(b)$, where $P_n(b)$ represents the power consumption of each phase with a b-bit resolution shifter \cite{MISO2018}. Therefore, the total power consumption of this system can be expressed as
\begin{align}\label{power_model3}
\mathcal{P}_{\rm total}= &\sum_{k=1}^{K}(\beta p_k + P_{{\rm U},k}) + P_{\rm B}+ NP_{n}(b).
\end{align}  \vspace{-2mm}

\subsection{Design Problem Formulation} \label{sec:problem}
We are interested in the joint design of the transmit rate for all users, included in $c=[c_1, \cdots, c_K]$, beamforming vector ${v}$, and the values for the RIS elements, appearing in the diagonal of $\mathbf{\Phi}=\mathrm{diag}[\phi_1,\phi_2,\ldots,\phi_N]$, that jointly maximize the bit-per-Joule EE performance for the considered RIS-based system. The base station release constant transmit power $P$ continuously.

The considered  maximize  power problem is expressed as follows:
\begin{subequations}\label{Prob:ResAllpower}
\begin{align}
\mathcal{P}_1: &\displaystyle \max_{\mathbf{\Phi},v_1,v_2,C_k} \frac{\sum^K_{k=1} C_k}{\sum_{k=1}^K  ||v_{1k}||^2_F + \sum_{k=1}^K  ||v_{2k}||^2_F + P} \label{Prob:aResAllpower}\\
\text{s.t.}
&\;\quad\;\; \sum_{k=1}^K  ||v_{1k}||^2_F + \sum_{k=1}^K  ||v_{2k}||^2_F \leq P_{max} \label{Prob:bResAllpower}\\
&\;  \quad\;\;\mathbf{C_k} \geq \mathbf{R};\label{Prob:cResAllpower}\\
&\;\quad\;\; |\phi_{1n}|=1,\;\forall n=1,2,\ldots,N, \label{Prob:dResAllpower}\\
&\;\quad\;\; |\phi_{2n}|=1,\;\forall n=1,2,\ldots,N, \label{Prob:eResAllpower}
\end{align}
\end{subequations}
where $C_k$ denotes the individual QoS constraint of the user, $P$ represents the total static power in the system. Constraint(3b) nsures that the sum of the powers to reach the RIS is equal to a fixed value. Constraint(3c) promises each user can get qualified signal whose SINR reached $\mathbf{R}$. Constraint(3d) and (3e) account for the fact that each RIS reflecting element can only provide a phase shift, without amplifying the incoming signal.

\section{Optimization Formulation And Analysis}\label{sec:format}
In this Section, we transform the non-convex problem into two sub-problems to solve transmitting beamforming optimization and the phase shifts. Firstly, for a given phase shift matrices, the corresponding optimization problem is transformed into an equivalent weighted minimum mean square error (WMMSE) problem formulation \cite{Cirik2013Weighted}. In the following, for the given solutions, we optimize the phase shift matrices via a gradient-based algorithm \cite{Wang2006Enhanced}. Repeat this cycle process until the convergence.
\subsection{ Transmit Beamforming Matrix Optimization with Fixed Phase Shift Matrix}
To deal with the complex objective function, we reformulate Problem P1 by employing the efficient $\mathbf{WMMSE}$ method \cite{Jian2016Queue}. Specifically, the linear decoding matrix $\mathbf{U}$ is applied to estimate the signal vector $\hat{s_k}$ for each user.$\hat{s}_{k} =\mathbf{U}_{k}^{H}y_{k} $, where $\mathbf{\mathbf{U}_k}\in \mathbb{C}^{M \times 1} $ is the decoding matrix of the $k$-th user.
Thus the MSE matrix \cite{Donoho2013The} of the $k$-th user is given by
\begin{equation*}
\begin{split}
&\mathbf{E_k} = E[|\hat{s_k}-s_k|^2] \\&= [\mathbf{U_k}^H(\overline{h}_{1k}v_{1k}+\overline{h}_{2k}v_{2k}) - \mathbf{I}][\mathbf{U_k}^H(\overline{h}_{1k}v_{1k}+\overline{h}_{2k}v_{2k}) - \mathbf{I}]^{H} \\&+\sum^K_{\bar{k}=1,\neq k}\mathbf{U_k}^H(\overline{h}_{1k}v_{1k}v_{1k}^H\overline{h}_{1k}^H +\overline{h}_{2k}v_{2k}v_{2k}^H\overline{h}_{2k}^H)\mathbf{U_k} + \sigma^2\mathbf{U_k}^H\mathbf{U_k}
\end{split}
\end{equation*}
where $s_k$ denote the data symbols of user $k$. By introducing additional variables $\mathbf{\mathbf{Q}_k}\in \mathbb{C}^{1 \times 1} $, we then have the following theorem:
\begin{subequations}\label{Prob:ResAllpower}
\begin{align}
\mathcal{P}_2: &\displaystyle \max_{\mathbf{Q}_k,\mathbf{U}_k,v_1,v_2} \sum^K_{k=1} [ln(\mathbf{Q_k}) - Tr(\mathbf{Q}_k\mathbf{E_k})]\label{Prob:aResAllpower}\\
\text{s.t.}
&\;\quad\;\; \sum_{k=1}^K  ||v_{1k}||^2_F + \sum_{k=1}^K  ||v_{2k}||^2_F \leq P \label{Prob:bResAllpower}\\
&\;  \quad\;\;ln(\mathbf{Q_k}) - Tr(\mathbf{Q}_k\mathbf{E_k}) \geq \mathbf{R}\label{Prob:cResAllpower}
\end{align}
\end{subequations}
By applying the standard convex optimization technique, setting the first-order derivative of the objective function of (P2) with respective to $\mathbf{U_k}$ and $\mathbf{Q_k}$ to zero, the optimal solutions of $\mathbf{U_k}$ and $\mathbf{Q_k}$ can be respectively obtained as
\begin{equation}
\mathbf{U_k}^{opt} = \mathbf{J} ^{-1}(\overline{h}_{1k}v_{1k}+\overline{h}_{2k}v_{2k})
\end{equation}
where $\mathbf{J} = \sum^K_{k=1}(\overline{h}_{1k}v_{1k}+\overline{h}_{2k}v_{2k})(\overline{h}_{1k}v_{1k}+\overline{h}_{2k}v_{2k})^{H} + \sigma^2\mathbf{I}$ is the covariance matrix of the total received signal at  $k$-th receiver. Via this MMSE receiver\cite{Jorswieck2003Transmission,Kim2008Performance}, the corresponding MSE matrix is given by
\begin{equation}
\mathbf{E_k}^{mmse} = \mathbf{I} - (\overline{h}_{1k}v_{1k}+\overline{h}_{2k}v_{2k})^H\mathbf{J}^{-1}(\overline{h}_{1k}v_{1k}+\overline{h}_{2k}v_{2k})
\end{equation}
By taking the first derivative of (5a)  and then set it equal to zero, we have 
\begin{equation}
\mathbf{Q_k}^{opt} = \mathbf{E_k}^{-1}
\end{equation}
Untill now the auxiliary matrix $\mathbf{U_k}$ and $\mathbf{Q_k}$ have been optimized, we focus on the optimization of transmitted power $v_{1k}$ and $v_{2k}$.
Accordingly we can turn P(2)  sum-utility maximization problem to the equivalent sum-MSE minimization problem\cite{Shi2007Downlink}. The update of transmit beamformers $v_{1k}$ and $v_{2k}$ for all can also be decoupled across transmitters, resulting in the following optimization problem:
\begin{subequations}\label{Prob:ResAllpower}
\begin{align}
\mathcal{P}_3: \displaystyle &\min_{v_1,v_2} \sum^K_{k=1} Tr(\mathbf{Q_k}[\mathbf{U_k}^H(\overline{h}_{1k}v_{1k}+\overline{h}_{2k}v_{2k})-I]\notag\\&[\mathbf{U_k}^H(\overline{h}_{1k}v_{1k}+\overline{h}_{2k}v_{2k})-I]^H) + \sum^K_{k=1}\sum^K_{j=1,\neq{k}}\notag\\&Tr[\mathbf{Q_k}\mathbf{U_k}^H(\overline{h}_{1k}v_{1j}v_{1j}^H\overline{h}_{1k}^H +\overline{h}_{2k}v_{2j}v_{2j}^H\overline{h}_{2k}^H)\mathbf{U_k}]\label{Prob:aResAllpower}\\
\text{s.t.}
&\;\sum_{k=1}^K  Tr(v_{1k}v_{1k}^H) + \sum_{k=1}^K Tr(v_{2k}v_{2k}^H) \leq P \label{Prob:bResAllpower}\\
&\;\quad\;\; Tr[\mathbf{Q_k}\mathbf{U_k}^H(\overline{h}_{1k}v_{1k}+\overline{h}_{2k}v_{2k})] + Tr\notag\\&[\mathbf{Q_k}(v_{1k}^H\overline{h}_{1k}^H + v_{2k}^H\overline{h}_{2k}^H)\mathbf{U_k}] -  \sum^K_{\bar{k}=1,\neq{k}}Tr\notag\\&[\mathbf{Q_k}\mathbf{U_k}^H(\overline{h}_{1k}v_{1j}v_{1j}^H\overline{h}_{1k}^H +\overline{h}_{2k}v_{2j}v_{2j}^H\overline{h}_{2k}^H)\mathbf{U_k}] \geq \mathbf{R} \label{Prob:cResAllpower}
\end{align}
\end{subequations}
This is a convex quadratic optimization problem which can be solved by using standard convex optimization algorithms\cite{Johansson2008Subgradient}. In fact, this problem also has a closed form solution using the Lagrange multipliers method. Specifically, attaching a Lagrange multiplier to the power budget constraint of transmitter, we get the following Lagrange function:
\begin{equation}
\begin{split}
&\Gamma(v_{1k},v_{2k},\mu_k) = \\&\sum^K_{k=1} \sum^K_{j = 1}Tr[\mathbf{Q_k}\mathbf{U_k}^H(\overline{h}_{1k}v_{1j}v_{1j}^H\overline{h}_{1k}^H +\overline{h}_{2k}v_{2j}v_{2j}^H\overline{h}_{2k}^H)\mathbf{U_k}] - \\&\sum^K_{k=1} Tr[\mathbf{Q_k}\mathbf{U_k}^H(\overline{h}_{1k}v_{1k}+\overline{h}_{2k}v_{2k})] - \sum^K_{k=1}Tr[\mathbf{Q_k}(v_{1k}^H\overline{h}_{1k}^H +\\&v_{2k}^H\overline{h}_{2k}^H)\mathbf{U_k}] + \mu_{k}(\sum_{k=1}^K  Tr(v_{1k}v_{1k}^H) + \sum_{k=1}^K Tr(v_{2k}v_{2k}^H) - P)
\end{split}
\end{equation}
The first-order optimized condition of $\Gamma(v_{1k},v_{2k},\mu_k)$ with respect to each $v_{1k},v_{2k}$ yields to
\begin{equation}
v_{1k}^{opt}(\mu_k) = (\sum^K_{k=1} \sum^K_{j = 1}\overline{h}_{1jk}^H\mathbf{U_k}\mathbf{Q_k}\mathbf{U_k}^H\overline{h}_{1jk} + \mu_k\mathbf{I})^{-1} \overline{h}_{1k}^H\mathbf{U_k}\mathbf{Q_k}
\end{equation}
\begin{equation}
v_{2k}^{opt}(\mu_k) = (\sum^K_{k=1} \sum^K_{j = 1}\overline{h}_{2jk}^H\mathbf{U_k}\mathbf{Q_k}\mathbf{U_k}^H\overline{h}_{2jk} + \mu_k\mathbf{I})^{-1} \overline{h}_{2k}^H\mathbf{U_k}\mathbf{Q_k}
\end{equation}
To obtain $v_{1k},v_{2k}$, the next thing we handle is to optimize $\mu_k$ to satisfy the complementarity slackness condition of the power budget constraint. We have
\begin{equation}
\sum_{k=1}^K  Tr(v_{1k}(\mu_k)v_{1k}^H(\mu_k)) + \sum_{k=1}^K Tr(v_{2k}(\mu_k)v_{2k}^H(\mu_k)) = P
\end{equation}
which is equivalent to 
\begin{equation}
Tr[(\Lambda_1 + \mu_k\mathbf{I})^{-2}\mathbf{M_1}] + Tr[(\Lambda_2 + \mu_k\mathbf{I})^{-2}\mathbf{M_2}] = P 
\end{equation}
where $T_1\Lambda_1T_1^H$,$T_2\Lambda_2T_2^H$ represent the eigendecomposition of $\sum^K_{k=1} \sum^K_{j = 1}\overline{h}_{1jk}^H\mathbf{U_k}\mathbf{Q_k}\mathbf{U_k}^H\overline{h}_{1jk}$ and $\sum^K_{k=1} \sum^K_{j = 1}\overline{h}_{2jk}^H\mathbf{U_k}\mathbf{Q_k}\mathbf{U_k}^H\overline{h}_{2jk}$ respectively. By matrix operation, $\mathbf{M_1}$, $\mathbf{M_2}$ express $\sum^K_{k=1}T_1^H(\overline{h}_{1k}^H\mathbf{U_k}\mathbf{Q_k}\mathbf{Q_k}^H\mathbf{U_k}^H\overline{h}_{1k})T_1$ and $\sum^K_{k=1}T_2^H(\overline{h}_{2k}^H\mathbf{U_k}\mathbf{Q_k}\mathbf{Q_k}^H\mathbf{U_k}^H\overline{h}_{2k})T_2$. Let $[\Lambda]_{mm}$ denote the mm-th diagonal element of $\Lambda$, then it can be simplified as:
\begin{equation}
\sum^M_{i = 1}\frac{[\mathbf{M_1}]_{mm}}{([\Lambda_1]_{mm} + \mu_k)^2} + \sum^M_{i = 1}\frac{[\mathbf{M_2}]_{mm}}{([\Lambda_2]_{mm}+ \mu_k)^2} = P
\end{equation}
We can find the optimal dual variable $\mu_k$ using bi-section search method based on the monotonic characteristic of the left side of (14). Moreover, to shrink the search space of the bi-section
method, we adopt the scaling method in the following upper bound
:
\begin{equation}
\begin{aligned}
\sum^M_{i = 1}\frac{[\mathbf{M_1}]_{mm}}{([\Lambda_1]_{mm} + \mu_k)^2} + \sum^M_{i = 1}\frac{[\mathbf{M_2}]_{mm}}{([\Lambda_2]_{mm}+ \mu_k)^2} \\\leq \sum^M_{i = 1}\frac{[\mathbf{M_1}]_{mm} + [\mathbf{M_2}]_{mm}}{(\mu_k^{max})^2} = P
\end{aligned}
\end{equation}
Moreover, $\mu_k^{max}$ can be simplified as:
\begin{equation}
\mu_k^{max} = \sqrt{\frac{\sum^M_{i = 1}[\mathbf{M_1}]_{mm} + \sum^M_{i = 1}[\mathbf{M_2}]_{mm}}{P}} 
\end{equation}
Above this equation, we only have to solve $\mu_k$ under known eigenmatrix. Note that the optimum $\mu_k$( denoted by $\mu_k^*$ ) must be positive in this case and the left-hand side of equation is a decreasing function in $\mu_k$ for $\mu_k > 0 $. So it is kinda easily solved using one dimensional search techniques. Finally, by plugging $\mu_k^*$, we get the solution for $v_{1k}(\mu_k^*)$ and  $v_{2k}(\mu_k^*)$.
So far we have figured out all variables in transmit beamforming matrix Optimization with fixed phase shift matrix. Next thing left for us is to iterate through the whole solution until target value converges under the SNR rate constraint (8c) .

\begin{algorithm}
	\caption{Subgradient Method for (P2)}
	\begin{algorithmic}[1]
		\State \textbf{Initialization:} $\mu_k^{n} \geq 0$,iteration index n = 0.$v_{1k}$ and $v_{2k}$ satisfy (8c).
		\State \textbf{REPEAT}
		\State  Solve the auxiliary matrix $\mathbf{U_k}$ and $\mathbf{Q_k}$ using (5) (7).
		\State  Calculate the optimal transmit beamforming matrix $V_1$, $V_2$ using (10),(11).
		\State  Compute dual variable $\mu_k^{n+1}$ using (16).
		\State  Set n = n + 1.
		\State \textbf{UNTIL} the fractional decrease of (9) is beyond a predefined threshold.
		\State \textbf{OUTPUT} $V_1^{opt}$ $V_2^{opt}$,$\forall n=1,2,\ldots,N,$
	\end{algorithmic}
\end{algorithm}

\subsection{Phase Shift Matrix Optimization with Fixed Transmit Beamforming}
In this subsection, we focus our attention on optimizing the phase shift matrix $\mathbf{\Phi_1}$ and $\mathbf{\Phi_2}$ while the other parameters are given. After the MSE method \cite{Donoho2013The} derivation and removing the terms that are independent of $\mathbf{\Phi}$, the phase shift optimization problem is formulated as:
\begin{subequations}\label{Prob:ResAllpower}
	\begin{align}
	\mathcal{P}_4: \displaystyle &\min_{\mathbf{\Phi_1},\mathbf{\Phi_2}} \sum^K_{k=1} \sum^K_{j = 1}Tr[\mathbf{Q_k}\mathbf{U_k}^H(\overline{h}_{1k}v_{1j}v_{1j}^H\overline{h}_{1k}^H +\notag\\&\overline{h}_{2k}v_{2j}v_{2j}^H\overline{h}_{2k}^H)\mathbf{U_k}] - \sum^K_{k=1} Tr[\mathbf{Q_k}\mathbf{U_k}^H(\overline{h}_{1k}v_{1k}+\overline{h}_{2k}v_{2k})] \notag\\& - \sum^K_{k=1}Tr[\mathbf{Q_k}(v_{1k}^H\overline{h}_{1k}^H +v_{2k}^H\overline{h}_{2k}^H)\mathbf{U_k}]\label{Prob:aResAllpower}\\
	\text{s.t.}
	&\;\quad\;\; Tr[\mathbf{Q_k}\mathbf{U_k}^H(\overline{h}_{1k}v_{1k}+\overline{h}_{2k}v_{2k})] + Tr[\mathbf{Q_k}(v_{1k}^H\overline{h}_{1k}^H \notag\\&+v_{2k}^H\overline{h}_{2k}^H)\mathbf{U_k}] -  \sum^K_{\bar{k}=1,\neq{k}}Tr[\mathbf{Q_k}\mathbf{U_k}^H(\overline{h}_{1k}v_{1j}v_{1j}^H\overline{h}_{1k}^H \notag\\&+\overline{h}_{2k}v_{2j}v_{2j}^H\overline{h}_{2k}^H)\mathbf{U_k}] \geq \mathbf{R} \label{Prob:bResAllpower}\\
	&\;\quad\;\; |\phi_{1n}|=1,\;\forall n=1,2,\ldots,N, \label{Prob:cResAllpower}\\
	&\;\quad\;\; |\phi_{2n}|=1,\;\forall n=1,2,\ldots,N, \label{Prob:dResAllpower}
	\end{align}
\end{subequations}
Let me restore the shorthand notation, substituting $\overline{h}_{1k} = g_{1k} \mathbf{\Phi_1}\mathbf{H_1}$ and $\overline{h}_{2k} = g_{2k} \mathbf{\Phi_2}\mathbf{H_2}$ back into P4. We have:
\begin{equation}
\begin{split}
&\mathbf{Q_k}\mathbf{U_k}^H(\overline{h}_{1k}v_{1j}v_{1j}^H\overline{h}_{1k}^H +\overline{h}_{2k}v_{2j}v_{2j}^H\overline{h}_{2k}^H)\mathbf{U_k} = \\&\mathbf{Q_k}\mathbf{U_k}^H(g_{1k} \mathbf{\Phi_1}\mathbf{H_1}v_{1j}v_{1j}^H\mathbf{H_1}^H\mathbf{\Phi_1}^Hg_{1k}^H +\\&g_{2k} \mathbf{\Phi_2}\mathbf{H_2}v_{2j}v_{2j}^H\mathbf{H_2}^H\mathbf{\Phi_2}^Hg_{2k}^H)\mathbf{U_k} 
\end{split}
\end{equation}
\begin{equation}
\begin{split}
&\mathbf{Q_k}\mathbf{U_k}^H(\overline{h}_{1k}v_{1k}+\overline{h}_{2k}v_{2k}) = \\&\mathbf{Q_k}\mathbf{U_k}^H(g_{1k} \mathbf{\Phi_1}\mathbf{H_1}v_{1k} + g_{2k} \mathbf{\Phi_2}\mathbf{H_2}v_{2k})
\end{split}
\end{equation}
Define $\mathbf{A_{1k}} = g_{1k}^H\mathbf{U_k}\mathbf{Q_k}\mathbf{U_k}^Hg_{1k}$, $\mathbf{A_{2k}} = g_{2k}^H\mathbf{U_k}\mathbf{Q_k}\mathbf{U_k}^Hg_{2k}$. $\mathbf{B_{1}} = \mathbf{H_1}\sum^K_{j = 1}(v_{1j}v_{1j}^H)\mathbf{H_1}^H$, $\mathbf{B_{2}} = \mathbf{H_2}\sum^K_{j = 1}(v_{2j}v_{2j}^H)\mathbf{H_2}^H$. Similarly, $\mathbf{C_{1k}} = \mathbf{H_1}v_{1k}\mathbf{Q_k}\mathbf{U_k}^Hg_{1k}$, $\mathbf{C_{2k}} = \mathbf{H_2}v_{2k}\mathbf{Q_k}\mathbf{U_k}^Hg_{2k}$.
By using this, we have:
\begin{equation}
\begin{split}
&tr(\mathbf{Q_k}\mathbf{U_k}^H(\overline{h}_{1k}v_{1j}v_{1j}^H\overline{h}_{1k}^H +\overline{h}_{2k}v_{2j}v_{2j}^H\overline{h}_{2k}^H)\mathbf{U_k}) = \\& tr(\mathbf{\Phi_1^H}\mathbf{A_{1k}}\mathbf{\Phi_1}\mathbf{B_{1}}) + tr(\mathbf{\Phi_2^H}\mathbf{A_{2k}}\mathbf{\Phi_2}\mathbf{B_{2}})
\end{split}
\end{equation}
\begin{equation}
tr(\mathbf{Q_k}\mathbf{U_k}^H(\overline{h}_{1k}v_{1k}+\overline{h}_{2k}v_{2k})) = tr(\mathbf{\Phi_1}\mathbf{C_{1k}}) + tr(\mathbf{\Phi_2}\mathbf{C_{2k}})
\end{equation}
By inserting (20) and (21) into the lines of Problem P4, we can get
\begin{subequations}\label{Prob:ResAllpower}
	\begin{align}
    \displaystyle &\min_{\mathbf{\Phi_1},\mathbf{\Phi_2}}   tr(\mathbf{\Phi_1^H}\mathbf{A_{1}}\mathbf{\Phi_1}\mathbf{B_{1}}) + tr(\mathbf{\Phi_2^H}\mathbf{A_{2}}\mathbf{\Phi_2}\mathbf{B_{2}}) \notag\\&-  tr(\mathbf{\Phi_1}\mathbf{C_{1}}) - tr(\mathbf{\Phi_2}\mathbf{C_{2}}) - tr(\mathbf{\Phi_1^H}\mathbf{C_{1}^H}) - tr(\mathbf{\Phi_2^H}\mathbf{C_{2}^H})\label{Prob:aResAllpower}\\
	 \text{s.t.}
	 &\;tr(\mathbf{\Phi_1}\mathbf{C_{1k}}) + tr(\mathbf{\Phi_2}\mathbf{C_{2k}}) + tr(\mathbf{\Phi_1^H}\mathbf{C_{1k}}^H) + tr(\mathbf{\Phi_2}^H\mathbf{C_{2k}}^H) \notag\\&- tr(\mathbf{\Phi_1^H}\mathbf{A_{1k}}\mathbf{\Phi_1}\mathbf{B_{1}}) - tr(\mathbf{\Phi_2^H}\mathbf{A_{2k}}\mathbf{\Phi_2}\mathbf{B_{2}}) \geq \mathbf{R} \label{Prob:bResAllpower}\\
	&\;\quad\;\; |\phi_{1n}|=1,\;\forall n=1,2,\ldots,N, \label{Prob:cResAllpower}\\
	&\;\quad\;\; |\phi_{2n}|=1,\;\forall n=1,2,\ldots,N, \label{Prob:dResAllpower}
	\end{align}
\end{subequations}
where $\mathbf{A_{1}}$ and $\mathbf{C_{1}}$ are given by $\mathbf{A_{1}} = \sum^K_{k=1} \mathbf{A_{1k}}$ and $\mathbf{C_{1}} = \sum^K_{k=1} \mathbf{C_{1k}}$, respectively.
Upon denoting the collection of diagonal elements of $\mathbf{\Phi}=\mathrm{diag}[\phi_1,\phi_2,\ldots,\phi_N]$, and adopting the matrix identity of SDR \cite{Luo2010Semidefinite,He2008Semidefinite}, it follows that
\begin{equation}
tr(\mathbf{\Phi^H}\mathbf{A}\mathbf{\Phi}\mathbf{B}) = \phi^H(\mathbf{A} \odot \mathbf{B^T})\phi
\end{equation}
Same as below, denoting the collections of diagonal elements of $\mathbf{C}$ by $c = [[\mathbf{C}]_{1,1}, · · · , [\mathbf{C}]_{N,N}]^T$ we arrive at
\begin{equation}
tr(\mathbf{\Phi}\mathbf{C}) = c^T\phi, tr(\mathbf{\Phi^H}\mathbf{C^H}) = \phi^Hc^*
\end{equation}
Thus the problem P4 can be transformed as
\begin{subequations}\label{Prob:ResAllpower}
	\begin{align}
	\displaystyle &\min_{\mathbf{\phi_1},\mathbf{\phi_2}} \phi_1^H\Psi_1\phi_1 + \phi_2^H\Psi_2\phi_2 - 2Re[\phi_1^Hc_1^*] - 2Re[\phi_2^Hc_2^*]\label{Prob:aResAllpower}\\
	&\; \text{s.t.}
	\; 2Re[\phi_1^Hc_{1k}^*] + 2Re[\phi_2^Hc_{2k}^*] - \phi_1^H\Psi_{1k}\phi_1 - \phi_2^H\Psi_{2k}\phi_2 \geq \mathbf{R}\label{Prob:bResAllpower}\\
	&\;\quad\;\; |\phi_{1n}|=1,\;\forall n=1,2,\ldots,N, \label{Prob:cResAllpower}\\
	&\;\quad\;\; |\phi_{2n}|=1,\;\forall n=1,2,\ldots,N, \label{Prob:dResAllpower}
	\end{align}
\end{subequations}
where we have $\Psi = \mathbf{A} \odot \mathbf{B^T}$. Moreover, $\mathbf{B}$ can be verified to be a non-negative semidefinite matrix, and thus $\Psi$ can be deduced as a non-negative semidefinite matrix.

However (25) is still hard to solve due to the non-convex constraint (25b). By using the SCA method\cite{2020Block} we can find this answer simply. Since $\phi^H\Psi\phi$ is convex, (25b) can be transformed as:
\begin{equation}
\begin{split}
&2Re[\phi_1^H(c_{1k}^* - \Psi_{1k}\phi_1^{(n)})] + 2Re[\phi_2^H(c_{2k}^* - \Psi_{2k}\phi_2^{(n)})] \geq\\& \mathbf{R} - \phi_1^{(n)H}\Psi_{1k}\phi_1^{(n)} - \phi_2^{(n)H}\Psi_{2k}\phi_2^{(n)} \triangleq \mathbf{\hat{R}}
\end{split}
\end{equation}
which is a linear constraint. Then, problem (25) can be written as:
\begin{subequations}\label{Prob:ResAllpower}
	\begin{align}
	\displaystyle &\min_{\mathbf{\phi_1},\mathbf{\phi_2}} \phi_1^H\Psi_1\phi_1 + \phi_2^H\Psi_2\phi_2 - 2Re[\phi_1^Hc_1^*] - 2Re[\phi_2^Hc_2^*]\label{Prob:aResAllpower}\\
	&\; \text{s.t.}
	\; 2Re[\phi_1^H(c_{1k}^* - \Psi_{1k}\phi_1^{(n)})] + 2Re[\phi_2^H(c_{2k}^* - \Psi_{2k}\phi_2^{(n)})] \geq \mathbf{\hat{R}}\label{Prob:bResAllpower}\\
	&\;\quad\;\; |\phi_{1n}|=1,\;\forall n=1,2,\ldots,N, \label{Prob:cResAllpower}\\
	&\;\quad\;\; |\phi_{2n}|=1,\;\forall n=1,2,\ldots,N, \label{Prob:dResAllpower}
	\end{align}
\end{subequations}
In the following we conceive the Majorization-Minimization (MM) algorithm \cite{Ying2016Majorization} for solving Problem (27). The key idea of using the MM algorithm lies in constructing a sequence of convex surrogate functions. Denoting the objective function of Problem (25) by $f(\phi)$. Specifically, at the n-th iteration, we need to construct an upper bound function of $f(\phi)$, denoted as $g(\phi|\phi^n)$, that satisfies the following three properties:
\begin{equation}
\begin{aligned}
&1). g(\phi^n|\phi^n)=f(\phi^n); \\ &2).\nabla_{\phi^n}g(\phi^n|\phi^n) = \nabla_{\phi^n}f(\phi^n);\\  &3)g(\phi|\phi^n) \geq f(\phi);
\end{aligned}
\end{equation}
where (1) denotes that $g(\phi|\phi^n)$ is an upper-bounded function of $f(\phi)$, (2) represents that $g(\phi|\phi^n)$ and $f(\phi)$ have the same solutions at point $\phi^n$, and (3) indicates $g(\phi|\phi^n)$ and $f(\phi)$ have the same gradient at point $\phi^n$.
Then, we solve the approximate subproblem defined by a more tractable new objective function $g(\phi|\phi^n)$.

Based on (28), at the n-th iteration:
\begin{equation}
\begin{aligned}
\phi^H\Psi\phi \leq \Lambda_{max}\phi^H\phi - 2Re[\phi^H(\Lambda_{max}\mathbf{I_N} - \Psi) \phi^n]\\+ \phi^{(n)H}(\Lambda_{max}\mathbf{I_N} - \Psi)\phi^n  \triangleq y(\phi|\phi^n)
\end{aligned}
\end{equation}
where $\Lambda_{max}$ is the maximum eigenvalue of $\Psi$. Therefore, at any r-th iteration, we define the objective function as
\begin{equation}
g(\phi|\phi^n) = y(\phi|\phi^n) - 2Re[\phi^Hc^*]
\end{equation}
Consequently our aim turns to be more tractable. Problem can be rewritten as:
\begin{subequations}\label{Prob:ResAllpower}
	\begin{align}
	\displaystyle &\min_{\mathbf{\phi_1},\mathbf{\phi_2}} g(\phi_1|\phi_1^n) + g(\phi_2|\phi_2^n)\label{Prob:aResAllpower}\\
	&\; \text{s.t.}
	\;2Re[\phi_1^H(c_1^* - \Psi_1\phi_1^{(n)})] + 2Re[\phi_2^H(c_2^* - \Psi_2\phi_2^{(n)})] \geq \mathbf{\hat{R}}\label{Prob:bResAllpower}\\
	&\;\quad\;\; |\phi_{1n}|=1,\;\forall n=1,2,\ldots,N, \label{Prob:cResAllpower}\\
	&\;\quad\;\; |\phi_{2n}|=1,\;\forall n=1,2,\ldots,N, \label{Prob:dResAllpower}
	\end{align}
\end{subequations}
We have $\phi^H\phi$ = N because of the eigenvalue characteristics. So $g(\phi|\phi^n)$ can be expressed as:
\begin{equation}
g(\phi|\phi^n) = \Lambda_{max}N - 2Re[\phi^Ht^n]
\end{equation}
where $t^n = c^* + (\Lambda_{max}\mathbf{I_N} - \Psi) \phi^n$.
So the problem can turn into:
\begin{subequations}\label{Prob:ResAllpower}
	\begin{align}
	\displaystyle &\max_{\mathbf{\phi_1},\mathbf{\phi_2}} 2Re[\phi_1^Ht_1^n] + 2Re[\phi_2^Ht_2^n]\label{Prob:aResAllpower}\\
	&\; \text{s.t.}
	\;2Re[\phi_1^H(c_{1k}^* - \Psi_{1k}\phi_1^{(n)})] + 2Re[\phi_2^H(c_{2k}^* - \Psi_{2k}\phi_2^{(n)})] \geq \mathbf{\hat{R}}\label{Prob:bResAllpower}\\
	&\;\quad\;\; |\phi_{1n}|=1,\;\forall n=1,2,\ldots,N, \label{Prob:cResAllpower}\\
	&\;\quad\;\; |\phi_{2n}|=1,\;\forall n=1,2,\ldots,N, \label{Prob:dResAllpower}
	\end{align}
\end{subequations}
Typically this kind of problem can be solved by using Lagrangian dual decomposition method. Introducing a non-negative $x$ on the left hand side of constraint (33a)
\begin{subequations}\label{Prob:ResAllpower}
	\begin{align}
	\displaystyle &\max_{\mathbf{\phi_1},\mathbf{\phi_2}} 2Re[\phi_1^Ht_1^n] + 2Re[\phi_2^Ht_2^n] + 2xRe[\phi_1^H(c_{1k}^* - \Psi_{1k}\phi_1^{(n)})] \notag\\&+ 2xRe[\phi_2^H(c_{2k}^* - \Psi_{2k}\phi_2^{(n)})]\label{Prob:aResAllpower}\\
	&\; \text{s.t.}
    \; |\phi_{1n}|=1,\;\forall n=1,2,\ldots,N, \label{Prob:bResAllpower}\\
	&\;\quad\;\; |\phi_{2n}|=1,\;\forall n=1,2,\ldots,N, \label{Prob:cResAllpower}
	\end{align}
\end{subequations}
Due to the independence between two variables. For a given $x$, the globally optimal solution is given by:
\begin{equation}
\phi_1(x) = e^{j(t_1^n + x(c_{1k}^* - \Psi_{1k}\phi_1^{(n)}))}
\end{equation}
\begin{equation}
\phi_2(x) = e^{j(t_2^n + x(c_{2k}^* - \Psi_{2k}\phi_2^{(n)}))}
\end{equation}
According to the monotony principle, the optimal value of $x$ should satisfy:
\begin{equation}
x(2Re[\phi_1^H(c_{1k}^* - \Psi_{1k}\phi_1^{(n)})] + 2Re[\phi_2^H(c_{2k}^* - \Psi_{2k}\phi_2^{(n)})] - \mathbf{\hat{R}}) = 0
\end{equation}
We discuss two scenarios to solve this:
1. $x$ = 0, then $\phi_1(0) = e^{j(t_1^n)}$ $\phi_2(0) = e^{j(t_1^n)}$ and have to satisfy constraint (26).
2. $x > 0$, equation (37) means $Y(x) = \mathbf{\hat{R}}$ where $Y(x) = 2Re[\phi_1(x)^H(c_{1k}^* - \Psi_{1k}\phi_1^{(n)})] + 2Re[\phi_2(x)^H(c_{2k}^* - \Psi_{2k}\phi_2^{(n)})]$.
It is not difficult to see $Y(x)$ is a monotonically increasing function of $x$. Based on the above discussions, we provide the algorithm to solve Problem (34) in
Algorithm 2.

\begin{algorithm}
	\caption{Bisection Search Method to solve constraint (8c)}
	\begin{algorithmic}[2]
		\State Calculate $Y(0)$. Make sure $Y(0) \geq \mathbf{\hat{R}}$.
		\State \textbf{Initialization:} accuracy $\epsilon$. lower bound $x_1$ and upper bound $x_2$.
		\State \textbf{REPEAT}
		\State  Calculate x = ($x_1$ + $x_2$)/2.
		\State  Update $\phi_1(x)$ and $\phi_2(x)$ using (35),(36) and Y(x).
		\State  If Y(x) $\geq \mathbf{\hat{R}}$, set $x_2 = x$, ohterwise $x_1 = x$.
		\State  If |$x_1 - x_2$| $\leq \epsilon$, break out the cycle.
	\end{algorithmic}
\end{algorithm}
Based on the above, we now provide the details of solving Problem (34) in Algorithm 3.
\begin{algorithm}
	\caption{MM Combined with SCA Algorithm to Solve Problem (34)}
	\begin{algorithmic}[3]
		\State \textbf{Initialization:} accuracy $\epsilon$,$\phi_1$,$\phi_2$,the iteration index to n = 0, the maximum number of iterations to $n_{max}$.
		\State \textbf{REPEAT}
		\State  Calculate $\mathbf{\hat{R}} = \mathbf{R} - \phi_1^{(n)H}\Psi_{1k}\phi_1^{(n)} - \phi_2^{(n)H}\Psi_{2k}
		\phi_2^{(n)}$ using (26).
		\State  Using $t^n = c^* + (\Lambda_{max}\mathbf{I_N} - \Psi) \phi^n$ to calculate $t_1^n$,$t_2^n$.
		\State  Update $\phi_1^{(n+1)}$ and $\phi_2^{(n+1)}$ using algorithm 2.
		\State  If $n \geq n_{max}$ or |$f(\phi^{(n+1)}) - f(\phi^{n})$| $\leq \epsilon$, break out the cycle.
	\end{algorithmic}
\end{algorithm}
Up to now we have figured out the B part phase shift matrix optimization. For given transmit beamforming matrix in section $\mathbf{A}$, we can solve the whole problem in algorithm 4.

\begin{algorithm}[!t]
	\caption{Alternating Algorithm for (P1).}
	\begin{algorithmic}[4]
		\State \textbf{Initialization:} iterative number n = 1, maximum number of iterations $n_{max}$, accuracy $\epsilon$ . $V_k^{(0)}$ and $\phi^{(0)}$.
		\State \textbf{REPEAT}
		\State  Calculate $\mathbf{U_1}^{opt}$,$\mathbf{U_2}^{opt}$ via algorithm 1.
		\State  Calculate $\mathbf{Q_1}^{opt}$,$\mathbf{Q_2}^{opt}$ via algorithm 1.
		\State  Calculate $\mathbf{V_1}^{opt}$,$\mathbf{V_2}^{opt}$ from algorithm 1.
		\State  Calculate $\Phi_1^{opt}$,$\Phi_2^{opt}$ from algorithm 3.
		\State \textbf{UNTIL} $n \geq n_{max}$or the fractional increase of P(1) is beyond a predefined threshold $\epsilon$.
	\end{algorithmic}
\end{algorithm}

\section{Numerical Results }\label{sec:format}

Here we provide numerical simulations to emulate the performance of Reconfigurable Intelligent Surfaces transmission system. The algorithm of this paper SDR compares to much different algorithms. We set up many other possible cases of the phase matrix for comparison. The other algorithms are respectively fixed phase shifts whose phase has been predetermined, all random phase shifts meaning all random phase on the RIS and same random phase shifts which selects the diagonal matrix of a random number.  Simulation results and analysis are shown in following:
\begin{figure} \vspace{-6mm}
		\begin{minipage}{9cm}
		\centering
		\includegraphics[width=6cm]{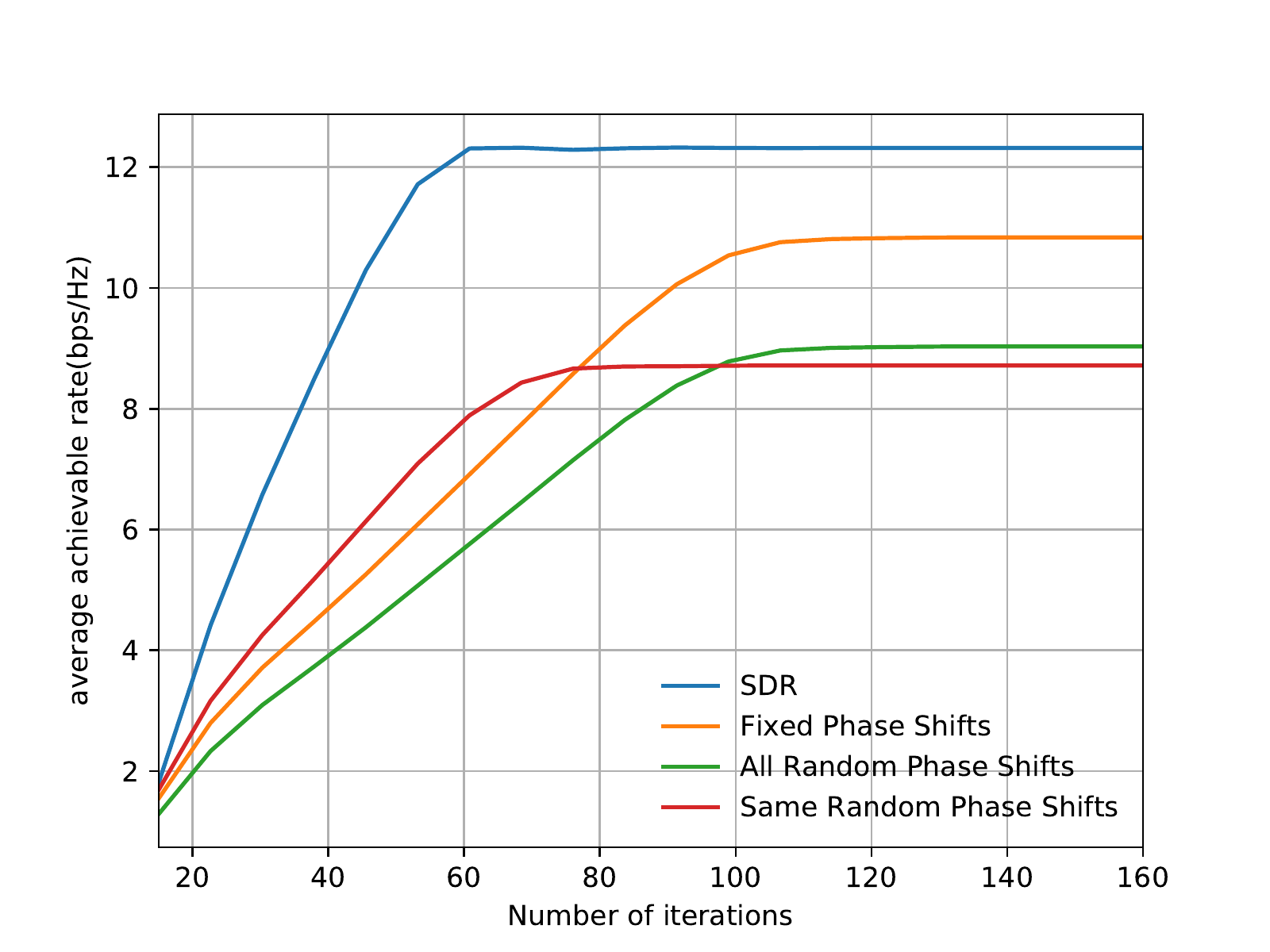}\vspace{-2mm}
		\caption{Comparison of algorithm performance while iteration number increases under same condition}
		\label{fig:Estimation_Scheme} \vspace{-0.5mm}
	\end{minipage}
	\begin{minipage}{9cm}
		\centering
		\includegraphics[width=6cm]{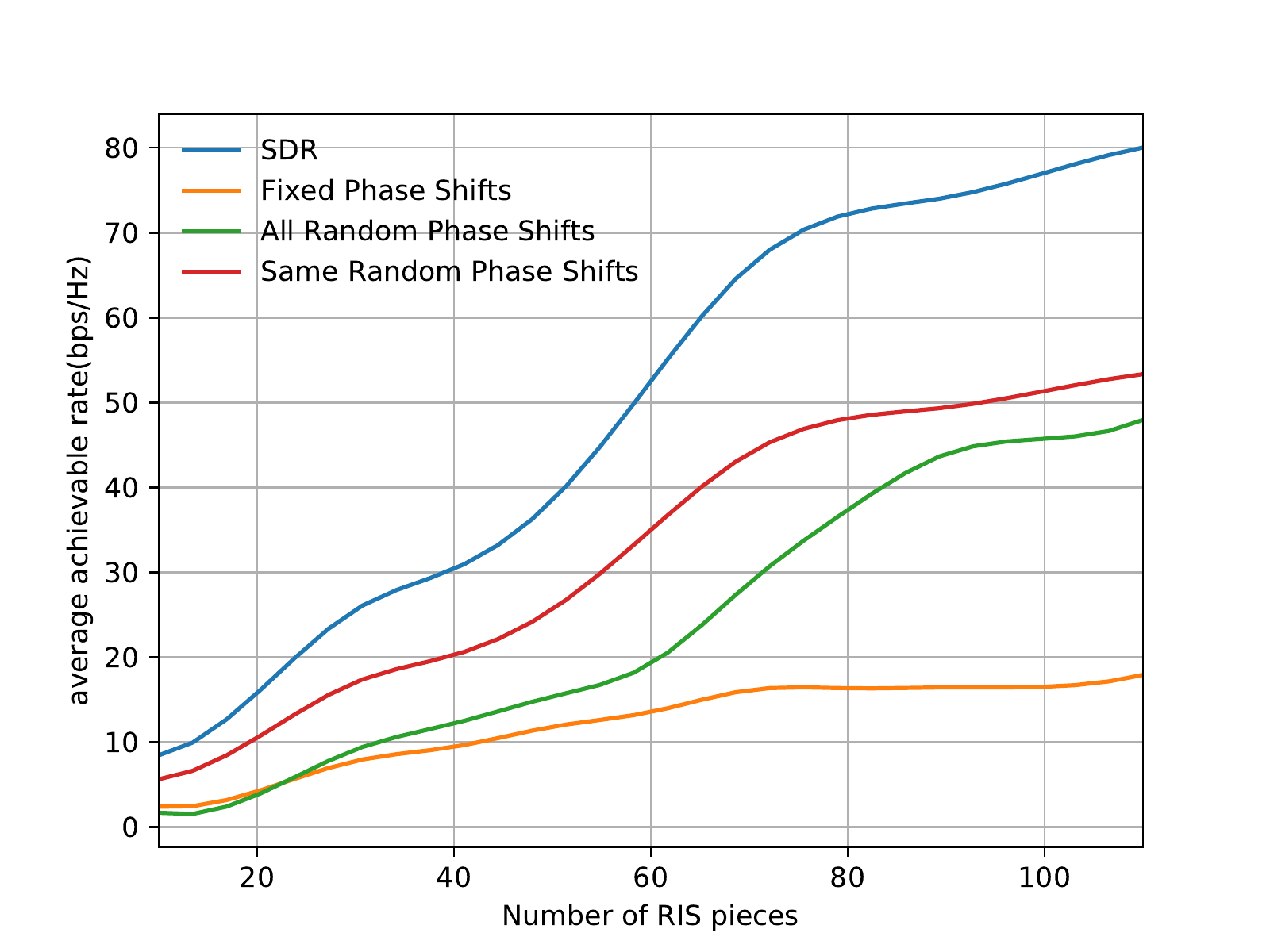}\vspace{-2mm}
		\caption{SDR algorithm performance with increasing RIS element numbers}
		\label{fig:Estimation_Scheme}
	\end{minipage}
\end{figure}

Figure 2 shows comparison of algorithm performance while iteration number increases under same condition whose Y-axis means average achievable rate. There we propose four scenarios: semidefinite relaxation (SDR), fixed phase shifts, all random phase shifts and same random phase shifts under circumstances where P = 60db, user K = 8, number of base station antennas M = 8, noise = 1, signal-to-noise ratio threshold R = 6.6 , RIS elements N = 16. As shown in figure 2, the total SNR increases with the increase of the number of cycles in every curve, and slowly arrive at their different balance point finally being constant. SDR algorithm performs higher than others and fastest come to its maximum. The other curves intersect when convergence is not reached, because the data are random.

We can conclude transmission efficiency is improved due to the increase of RIS element numbers from Figure 3 under circumstances where P = 60db, user K = 8, number of base station antennas M = 8, noise = 1, signal-to-noise ratio threshold R = 6.6 and cycle index 200 satisfying convergence. In the given uniform other conditions meet the transmission conditions, SDR algorithm always perform better than other algorithms. The more number of RIS pieces, the larger RIS area will be. So it can reflect more transmitting signals and reduce leakage. Thus improve the average achievable rate. The increasing curves grow fast or slow because area is to the second power not linear.

\begin{figure} \vspace{-1mm}
	\begin{minipage}{9cm}
		\centering
		\includegraphics[width=6cm]{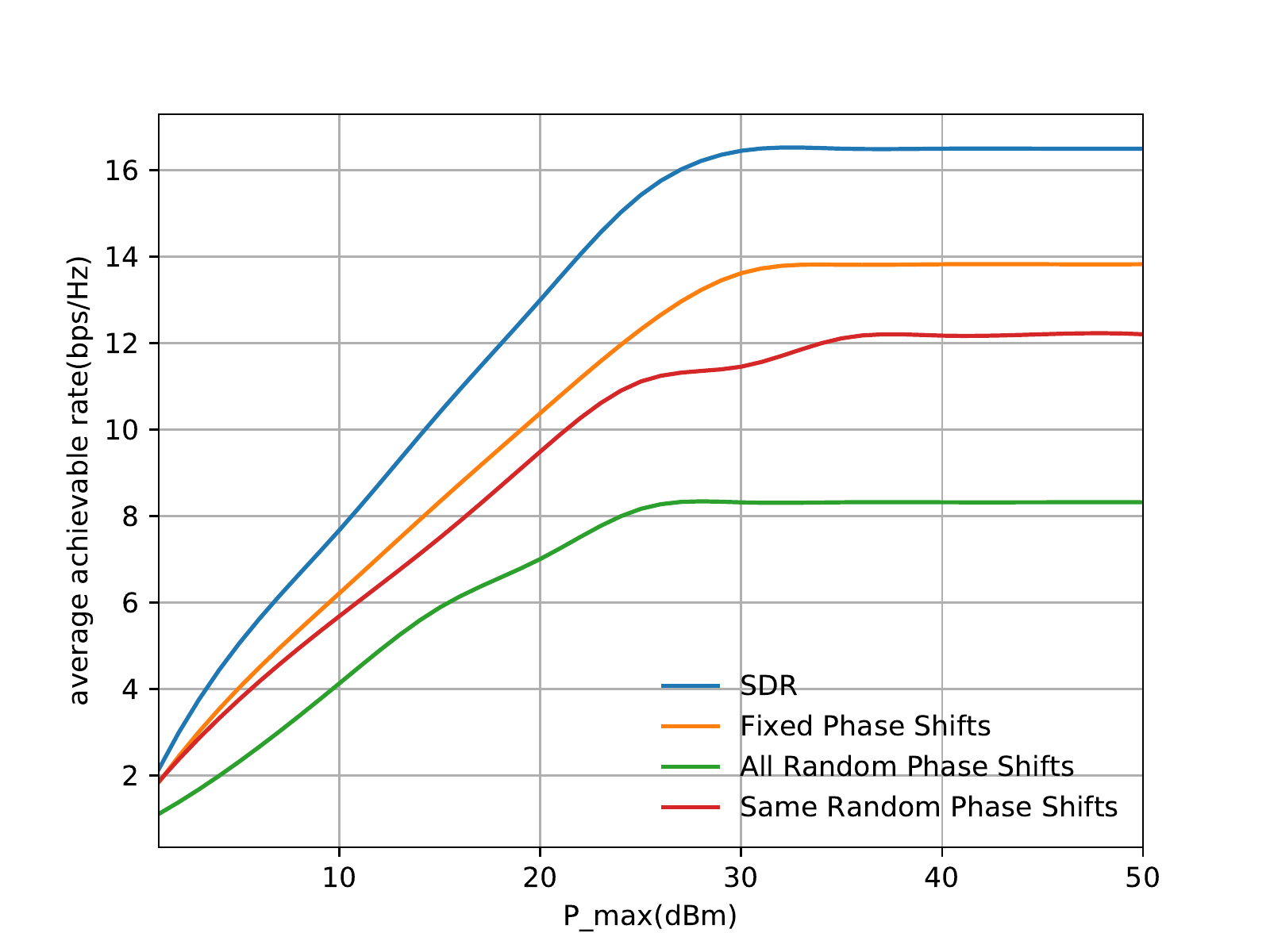}\vspace{-2mm}
		\caption{Comparison of algorithm performance when P-max as the power from the base station increases}
		\label{fig:Estimation_Scheme} \vspace{-0.5mm}
	\end{minipage}
	\begin{minipage}{9cm}
		\centering
		\includegraphics[width=6cm]{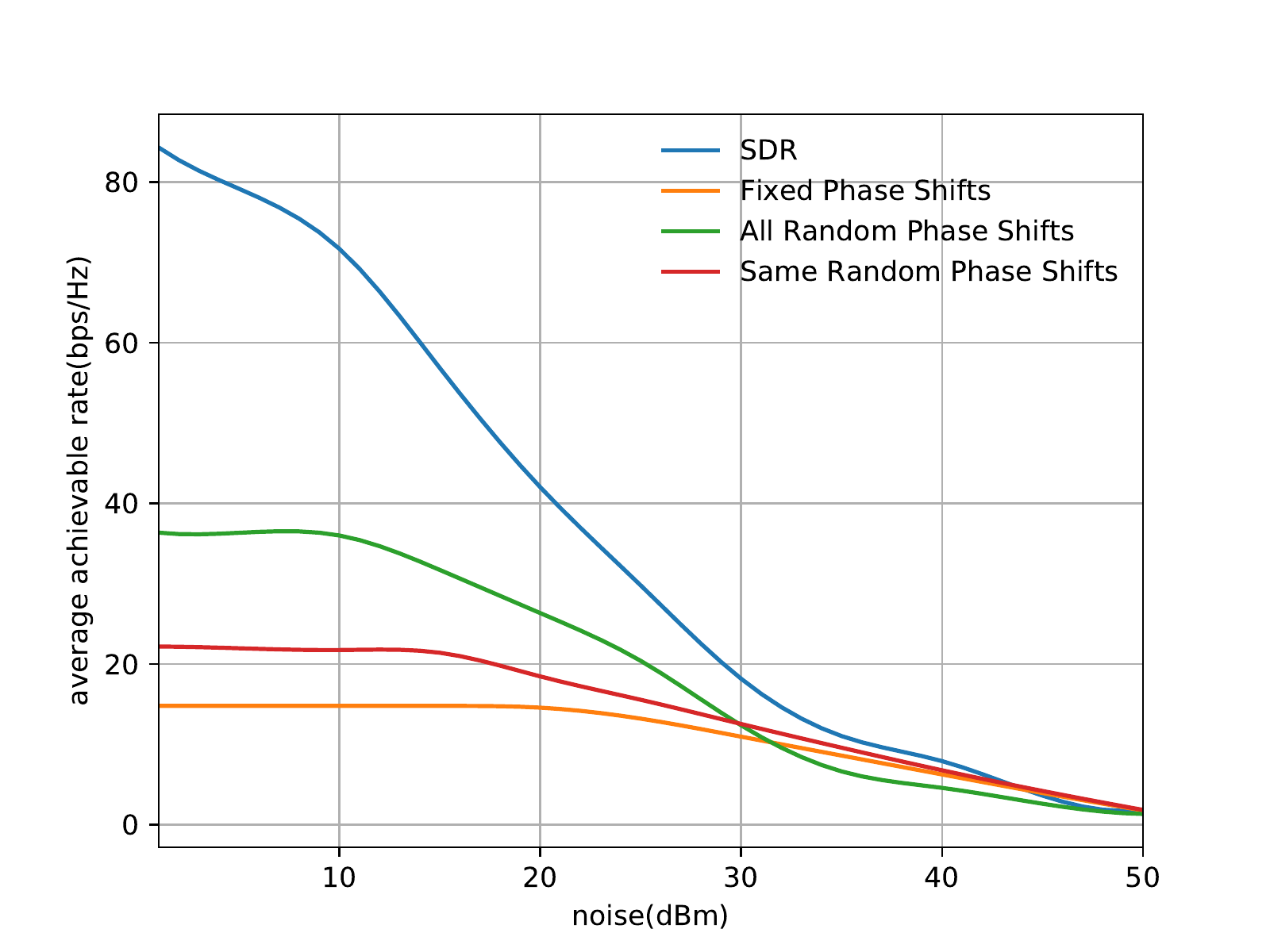}\vspace{-2mm}
		\caption{Comparison of performance of each algorithm when adjusting system noise}
		\label{fig:Estimation_Scheme}
	\end{minipage}
\end{figure}

From Figure 4 we found that total signal-to-noise ratio grows with max energy power emitted by the base station, which means when P-max is not big enough there's plenty room in the system channel. This set of data is produced under this condition where user K = 8, number of base station antennas M = 8, noise = 1, signal-to-noise ratio threshold R = 6.6 , RIS elements N = 16 and cycle index 200 satisfying convergence. Meanwhile when P-max retains a threshold, average achievable rate comes to a plain reaching the maximum limit of the system. SDR algorithm curve always higher than others because the coupling optimization of transmitting matrix and phase shifts.

After the positive comparison, performance of different systems in the face of increased noise is also documented in Figure 5 under situation where P = 60db, user K = 8, number of base station antennas M = 8, RIS elements N = 16, signal-to-noise ratio threshold R = 6.6 and cycle index 200 satisfying convergence..
Viewing from left to right on the horizontal axis, as the noise increases slightly, the signal-to-noise rate curves decline mildly. However the overall downward trend of the entire curve is inevitable due to increasing noise sound. As the noise gradually increases in the middle of the X-axis, the SNR curve decreases rapidly. The advantage of SDR algorithm shows in that even when they are all going down it's still on top of other curves. Finally the noise is too large and the SNR gradually approaches zero where extreme situation is simulated which rarely appears in real life just for theoretical research.
\begin{figure} \vspace{-1mm}
	\begin{center}
		\includegraphics[width=6cm]{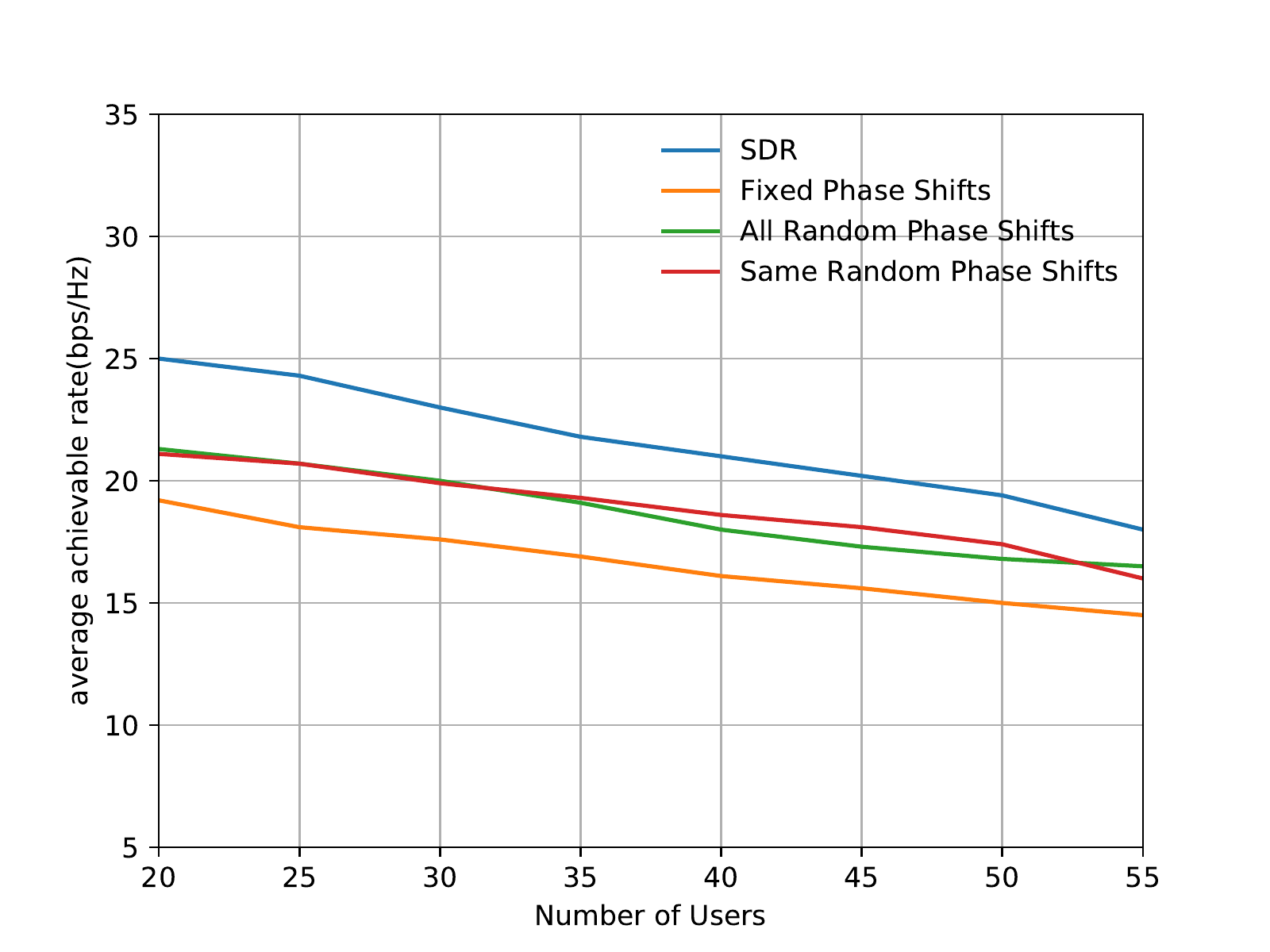}  \vspace{-2mm}
		\caption{Comparison of algorithm performance when user numbers are increasing}
		\label{fig:Estimation_Scheme} \vspace{-6mm}
	\end{center}
\end{figure}

Figure 6 shows comparison of algorithm performance when user numbers are increasing under circumstances where P = 60db, number of base station antennas M = 8, noise = 1, RIS elements N = 16, signal-to-noise ratio threshold R = 6.6 and cycle index 200 satisfying convergence.. More users means more channel frequency usage, which also means more energy consumption and more interfering noise. We can learn SDR algorithm curve is higher than others while all total average achievable rate curves are declining. Their tendency is almost linear due to the linear user growth.

\section{Conclusion And Future Work}\label{sec:format}

In this paper, we designed a Reconfigurable Intelligent Surface (RIS) for downlink multi-user communication system. In order to accomplish better power transmission performance, we focus on maximizing the total SNR of receiving users issue by jointly optimizing the two different RIS phase shift matrix and the BS transmit beamforming matrix. It is more efficient to transmit signals via two opposite RIS. According to different transfer matrices between BS(base station) and RIS, we solve the coupling optimization using semidefinite relaxation techniques to simplify transmitting beamforming matrix and the matrix phase shift. After the two problems have been solved separately, an efficient iterative algorithm is adopted for integrated treatment. Simulation results demonstrated that the algorithm performed better than other comparison examples, which means it did improve energy collection performance. This paper assumes known channel parameter performance, which may be hard to get in reality. For the future work, we will consider RIS system where CSI is hypothesised to be imperfectly known. Additionally, as the distance of transmission changes, the electromagnetic transmission characteristic of RIS may be altered by geographical distance from base station or from users.

\bibliographystyle{IEEEtran}
\bibliography{reference}
	
\end{document}